\documentclass[submission, Phys]{SciPost}

\hypersetup{
    colorlinks,
    linkcolor={red!50!black},
    citecolor={blue!50!black},
    urlcolor={blue!80!black}
}

\usepackage[bitstream-charter]{mathdesign}
\DeclareSymbolFont{usualmathcal}{OMS}{cmsy}{m}{n}
\DeclareSymbolFontAlphabet{\mathcal}{usualmathcal}

\newcommand{\mcC}{\mathcal{C}}
\newcommand{\dd}{\mathrm{d}}
\newcommand{\ee}{\boldsymbol{e}}
\newcommand{\EE}{\boldsymbol{E}}
\newcommand{\ff}{\boldsymbol{f}}
\newcommand{\jj}{\boldsymbol{j}}
\newcommand{\JJ}{\boldsymbol{J}}
\newcommand{\kk}{\boldsymbol{k}}
\newcommand{\mcO}{\mathcal{O}}
\renewcommand{\ss}{\boldsymbol{s}}
\newcommand{\uu}{\boldsymbol{u}}
\newcommand{\ww}{\boldsymbol{w}}
\newcommand{\xx}{\boldsymbol{x}}

\newcommand{\nnu}{\boldsymbol{\nu}}
\newcommand{\zzeta}{\boldsymbol{\zeta}}

\newcommand{\ind}[1]{_{\mathrm{#1}}}

\newcommand{\ldeb}{\lambda\ind{D}}

\newcommand{\transp}{^\mathrm{T}}

\DeclareMathOperator{\erfc}{erfc}

\begin{document}


\begin{center}
{\Large \textbf{Temporal response of the conductivity of electrolytes}}
\end{center}

\begin{center}
Haggai Bonneau\textsuperscript{1},
Vincent Démery\textsuperscript{1,2$\star$} and
Elie Raphaël\textsuperscript{1}
\end{center}

\begin{center}
{\bf 1} Gulliver, UMR CNRS 7083, ESPCI Paris PSL, 75005 Paris, France
\\
{\bf 2} Université Lyon, ENS de Lyon, Université Claude Bernard, CNRS, Laboratoire de Physique, F-69342 Lyon, France
\\
${}^\star$ {\small \sf vincent.demery@espci.psl.eu}
\end{center}

\begin{center}
\today
\end{center}


\section*{Abstract}
{\bf
We study the temporal response of the electric current in an electrolyte under a sudden switch on or switch off of an external electric field of arbitrary magnitude.
We use Stochastic Density Functional Theory including hydrodynamic interactions to express the current as a function of the ionic correlations.
Assuming small density fluctuations, we linearize the field theory to compute the correlations in the transient regime.
We first show that the correlations do not follow the same trajectory when the field is switched on or switched off.
Accordingly, the behavior of the current differs in the two cases:
it decays exponentially when the field is switched off, but it relaxes algebraically to its stationnary value when the field is switched on.
This difference is a non-linear effect since an exponential relaxation is recovered in both cases in the weak field limit.
}

\vspace{10pt}
\noindent\rule{\textwidth}{1pt}
\tableofcontents\thispagestyle{fancy}
\noindent\rule{\textwidth}{1pt}
\vspace{10pt}

\section{Introduction}

Determining the response of an electrolyte to an external field is a century-old non-equilibrium statistical physics problem.
Its theoretical exploration started with the works of Debye and Hückel~\cite{Debye1923}, and later of Onsager~\cite{Onsager1927}, who settled the so-called DHO theory for the weak field conductivity of electrolyte solutions.
The finite field solution was provided thirty years later by Onsager and Kim~\cite{Onsager1957Wien}.
The DHO theory contains two corrections to the naive conductivity, which arise from the correlations between ions: the first, which we call the electrostatic correction, comes from the electric field generated by the disturbed cloud of counterions surrounding a given ion. 
The second, which we call the hydrodynamic correction, takes into account the hydrodynamic flow generated by the cloud of counterions.

The study of the conductivity of electrolytes has been revived recently.
First, Stochastic Density Field Theory (SDFT) has been used to compute the ionic correlations in absence of hydrodynamic interactions, giving access to the electrostatic correction to the conductivity~\cite{Dean1996, Demery2016Conductivity}.
Then, SDFT has been coupled with fluctuating hydrodynamics, providing a stochastic field theory for the ionic densities and the flow, allowing to derive the electrostatic correction together with the hydrodynamic one~\cite{Peraud2017,Donev2019}.
These more systematic approaches allowed further progress, such as taking into account the finite size of the ions using a slight modification of the interaction kernel, leading to quantitative predictions up to large densities~\cite{Avni2022Conductivity,Avni2022Conductance}, or unveiling long range forces between objects immersed in a driven electrolyte~\cite{Mahdisoltani2021}.
In parallel, the effect of the correlations on the conductivity has been studied through numerical simulations, on lattice~\cite{Kaiser2013} or using molecular dynamics~\cite{Lesnicki2020,Lesnicki2021}, with an implicit or explicit solvent.

The above mentionned works addressed the non-equilibrium stationnary state (NESS) of a driven electrolyte.
Under a time-dependent driving, new interesting effects appear, such as a long-ranged repulsion between oppositely charged surfaces under a periodic driving~\cite{Richter2020}, or synapse-like memory effects in strongly confined electrolytes~ \cite{Robin2021Modeling,Robin2023Long-term}, and ``magnetolytes'' in spin ice become formally equivalent to an electrolyte~\cite{Kaiser2015AC}.
To date, there are little examples where the field theoretic machinery described above has been used to study transient effects in electrolytes.
One is the calculation of the transient fluctuations induced forces between two objects immersed in a driven electrolyte upon a sudden field change~\cite{Mahdisoltani2021Transient}; however, the hydrodynamic interactions have been neglected and the calculation of transient correlations limited to distances much larger than the Debye length.
Other related examples are the calculations of the relaxation of the Casimir force between two polarizable slabs or two conducting plates~\cite{Dean2012, Dean2014}, but in those cases the transition takes place between two equilibrium states.

Here, we use SDFT with hydrodynamic interactions~\cite{Demery2016Conductivity,Donev2019} to study the evolution of the electric current in a bulk electrolyte when the external field is suddenly switched on or off.
First, we show that the correlations do not follow the same path when the system goes from equilibrium to NESS, or from NESS to equilibrium.
Second, we turn to the current, and in particular to the electrostatic and hydrodynamic corrections.
From NESS to equilibrium, the hydrodynamic correction is absent and we find that the electrostatic correction decays exponentially.
On the contrary, from equilibrium to NESS, we unveil an algebraic decay of both corrections.
At linear order in the field, the corrections decay exponentially, showing that the algebraic decay is a non-linear effect.
Finally, we discuss the relationship between the relaxations of the correlation and of the current.

This article is organized as follows.
The model is introduced in Sec.~\ref{sec:model}.
In Sec.~\ref{sec:correlations_current} we obtain a closed equation for the density fields of the ions, express the electric current as a function of the correlations of the density fields, and derive and solve the equation for the correlations in the transient regime, assuming Gaussian density fields.
We compute and analyse the corrections from NESS to equilibrium in Sec.~\ref{sec:ness_eq}, and from equilibrium to NESS in Sec.~\ref{sec:eq_ness}. 
We conclude in Sec.~\ref{sec:conclusion}.

\section{Model}
\label{sec:model}

We consider a system of charged Brownian particles of different species in a three dimensional homogeneous solution, subjected to a uniform external electric field with a time dependent amplitude $\EE(t)=E(t)\hat\ee_x$, where $\hat\ee_x$ is the unit vector along the $x$-axis. 
The particles interact via the electrostatic potential and are advected by the flow in the solution, which is generated by the forces transmitted by the particles on the solvent.
We denote $\bar\rho_\alpha$ the average density of the particles of the species $\alpha$, $\kappa_\alpha$ their mobility and $qz_\alpha$ their charge, with $q$ being the elementary charge.
We assume that the system is electroneutral: $\sum_\alpha z_\alpha\bar\rho_\alpha=0$.

We describe the evolution of the density field $\rho_\alpha(\xx,t)$ of the species $\alpha$ using Stochastic Density Functional Theory~\cite{Dean1996,Demery2016Conductivity} with hydrodynamic interactions~\cite{Peraud2017,Donev2019}:
\begin{align}
\dot\rho_\alpha &= - \nabla\cdot \boldsymbol{j}_\alpha, \label{eq:continuity}\\
\jj_\alpha
	&= \boldsymbol{u}\rho_\alpha -T\kappa_\alpha \nabla \rho_\alpha
	+ \kappa_\alpha   \rho_\alpha \boldsymbol{f}_\alpha
	+\sqrt{\kappa_\alpha T \rho_\alpha} \boldsymbol{\zeta}_\alpha \label{eq:particle_current}
\end{align}
where $\boldsymbol{u}(\xx,t)$ is the velocity field of the solution, $T$ is the temperature (we set the Boltzmann constant to $k\ind{B}=1$) and $ \boldsymbol{f}_\alpha(\xx,t)$ is the force acting on the particles of the species $\alpha$.
The noise term $\boldsymbol{\zeta}(\xx,t)$ is a Gaussian white noise with the correlation
\begin{equation}\label{eq:noise}
\langle \boldsymbol{\zeta}_\alpha(\xx,t) \boldsymbol{\zeta}_\beta(\xx',t') \rangle
	= 2 \delta_{\alpha\beta} \delta(\xx-\xx')\delta(t-t').
\end{equation}
We use the Itô convention for the multiplicative noise in Eq.~(\ref{eq:particle_current}) and throughout the manuscript~\cite{Oksendal2000,Dean1996}.

The force on the particles of the species $\alpha$ is the sum of the force exerted by the external field and the force due to pair interactions:
\begin{equation}
\ff_\alpha=z_\alpha q \boldsymbol{E} - \sum_\beta \nabla V_{\alpha\beta}*\rho_\beta,
\end{equation}
where $V_{\alpha\beta}(\xx)=q^2 z_\alpha z_\beta/(4\pi\epsilon r)$ is the electrostatic interaction, with $r=|\xx|$, $\epsilon$ the dielectric permittivity of the solvent, and $*$ the convolution operator.

We assume that the fluid velocity field $\uu(\xx,t)$ satisfies the fluctuating Stokes equation for incompressible fluids~\cite{DeZarate2006Hydrodynamic} (Sec.~3.2):
\begin{align}
\nabla\cdot \boldsymbol{u} & = 0 \label{eq:incompress}\\
-\eta\nabla^2 \boldsymbol{u} - \nabla p &= \sum_\alpha \rho_\alpha \ff_\alpha+\sqrt{\eta T} \nabla\cdot \left(\nnu +\nnu\transp\right) \label{eq:stokes}
\end{align}
where $\nnu(\xx,t)$ is a Gaussian noise tensor field with correlation function:
\begin{equation}
\langle \nu_{ij}(\xx,t) \nu_{kl}(\xx',t') \rangle=\delta_{ik}\delta_{jl}\delta(\xx-\xx')\delta(t-t').
\end{equation}

We compute the total average electric current $\JJ(t)$,
\begin{equation}\label{eq:average_electric_current}
\boldsymbol{J}=q\sum_\alpha z_\alpha \langle \boldsymbol{j}_\alpha \rangle= J \hat\ee_x,
\end{equation} 
and then discuss the correction to the current without interactions, $\sigma_0\EE$, 
where $\sigma_0 = q^2\sum_\alpha z_\alpha^2 \kappa_\alpha \bar\rho_\alpha$ is the bare conductivity of the solution.
In particular, we are interested in the evolution of the current $J(t)$ when the electric field is suddenly switched on ($E(t)=E_0 H(t)$, where $H(t)$ is the Heaviside function), or off ($E(t)=E_0 H(-t)$).
In the first case, the system goes from equilibrium with $E=0$ to a non-equilibrium steady state (NESS) with $E=E_0$; in the second case, the system relaxes from a NESS to equilibrium.

\section{Correlations and electric current}
\label{sec:correlations_current}

\subsection{Closed equations for the density fields}
\label{sub:closed_density}

We can integrate the fluid degrees of freedom $\uu$ to obtain a closed equation for the densities $\rho_\alpha$.
The solution to Eqs.~(\ref{eq:incompress}, \ref{eq:stokes}) is given by the convolution of the force density (the right hand side of Eq.~(\ref{eq:stokes})) with the Oseen tensor, $\mathcal{O}_{ij}(\xx) = \frac{1}{8\pi \eta }\left( \frac{\delta_{ij}}{r}+\frac{x_i x_j}{r^3}\right)$~\cite{Kim2013} (Chap.~2).
Inserting this result in the expression for the density current, Eq.~(\ref{eq:particle_current}), we get
\begin{equation}
\boldsymbol{j}_{\alpha}= -\kappa_{\alpha}T\nabla \rho_{\alpha} +\kappa_{\alpha}\rho_\alpha \boldsymbol{f}_{\alpha} +\rho_{\alpha} \sum_\beta \mathcal{O}*\left[  \rho_\beta\boldsymbol{f}_\beta\right] 
+\sqrt{\kappa_{\alpha}T\rho_{\alpha}} \boldsymbol{\zeta}_{\alpha}+\sqrt{\eta T} \rho_\alpha\boldsymbol{w}
\label{eq:current_expanded}
\end{equation}
where we have introduced the Gaussian noise vector field $\ww(\xx,t)$ with correlation
\begin{equation}\label{eq:noise_oseen}
\left\langle w_i (\xx,t)w_j (\xx',t')\right\rangle =
2 {\mathcal{O}}_{ij}(\xx-\xx') \delta(t-t').
\end{equation}
Equations (\ref{eq:continuity}, \ref{eq:noise}, \ref{eq:current_expanded}, \ref{eq:noise_oseen}) form a closed set of equations for the densities.

Our procedure is however not completely correct: when a force is applied on a particle, it gives rise to a flow that is given by the Oseen tensor.
However, this flow diverges at the location of the particle, giving the particle  an infinite velocity~\cite{Brogioli2000Diffusive}.
Moreover, the motion of the particle resulting from the application of the force is already taken into account by the mobility of the particle.
Hence, when computing the flow advecting a given particle, one should take care to omit the flow created by the forces acting on this particle.
There is no simple way to do it in our field theory, but this flaw is easily corrected when the electric current is expressed with the correlations.

\subsection{Average electric current from correlations}%

Using the expression (\ref{eq:current_expanded}) in the average electric current (Eq.~(\ref{eq:average_electric_current})) leads to
\begin{equation}\label{eq:average_electric_current2}
\JJ=q \sum_\alpha z_\alpha \left\langle \kappa_{\alpha}\rho_\alpha \boldsymbol{f}_{\alpha}+\rho_{\alpha} \sum_\beta\mathcal{O}*\left[\rho_\beta\boldsymbol{f}_\beta\right]\right\rangle.
\end{equation}
Note that the noise terms cancel as they are uncorrelated to the density fields and the gradient term cancels as we assume spatial invariance.

We now express the average electric current as a function of the correlations of the density fields.
We introduce the density fluctuations $n_\alpha(\xx,t)$,
\begin{equation}
\rho_\alpha(\xx,t) = \bar\rho_\alpha +n_\alpha(\xx,t),
\end{equation}
and the correlation
\begin{equation}
C_{\alpha\beta}(\xx-\xx',t) =\langle n_\alpha(\xx,t) n_\beta(\xx',t)\rangle=  \mathcal{C}_{\alpha\beta}(\xx-\xx',t)+\bar{\rho}_\alpha\delta_{\alpha\beta}\delta(\xx-\xx').
\label{eq:def_correlation}
\end{equation}
where $\mathcal{C}$ is the pair correlation function, which does not contain the self correlation.

Using electroneutrality, the average electric current (Eq.~(\ref{eq:average_electric_current2})) can be expressed with the density fluctuations,
\begin{multline}
\JJ = \sigma_0 \EE - \sum_{\alpha,\beta} qz_\alpha\kappa_\alpha \langle n_\alpha \nabla V_{\alpha\beta}*n_\beta\rangle 
+ \sum_{\alpha,\beta} q^2z_\alpha z_\beta \langle n_\alpha\mcO*n_\beta \rangle\EE 
\\- \sum_{\alpha,\beta,\gamma} qz_\alpha \langle n_\alpha \left[\mcO*(n_\beta[\nabla V_{\beta\gamma}*n_\gamma]) \right] \rangle.
\end{multline}
Writing the convolutions explicitly and using the correlation (Eq.~(\ref{eq:def_correlation})), we arrive at
\begin{multline}\label{eq:current_correlations}
\JJ = \sigma_0 \EE - \sum_{\alpha,\beta} qz_\alpha\kappa_\alpha \int\nabla V_{\alpha\beta}(\xx)C_{\alpha\beta}(\xx)\dd\xx
+\sum_{\alpha,\beta} q^2z_\alpha z_\beta \int \mcO(\xx)C_{\alpha\beta}(\xx)\dd\xx \,\EE 
\\-\sum_{\alpha,\beta,\gamma}qz_\alpha \int \mcO(\xx)\nabla V_{\beta\gamma}(\xx')C^{(3)}_{\alpha\beta\gamma}(\xx,\xx')\dd\xx\dd\xx',
\end{multline}
where we have introduced the three-point correlation $C^{(3)}_{\alpha\beta\gamma}(\xx-\xx',\xx'-\xx'')=\langle n_\alpha(\xx) n_\beta(\xx') n_\gamma(\xx'') \rangle$.

The correction to the bare current $\sigma_0\EE$ is the sum of three contributions:
\begin{itemize}
\item The first involves the correlation and the electrostatic potential, we call it the \emph{electrostatic correction} (it was originally called the relaxation correction).
It represents the effect of the electric field of the cloud of counterions around a charged particle, which is deformed when an external field is applied.
\item The second term involves the correlation, the Oseen tensor, and the external field, we call it the \emph{hydrodynamic correction} (it was originally called the electrophoretic correction).
It contains the effect of the flow created by the cloud of counterions under the action of the external field.
\item The last term combines electrostatic and hydrodynamic effects: it contains the effect of the flow created by the counterions under the action of electrostatic interactions between the particles. 
As electrostatic and hydrodynamic interactions are involved, their interaction kernels are coupled to the three-point correlation $C^{(3)}$.
\end{itemize}

At this stage, we can make the correction mentionned in Sec.~\ref{sub:closed_density}: remove the effect of the flow that is generated by a particle on this same particle.
In the hydrodynamic correction, this is done by replacing the correlation $C_{\alpha\beta}(\xx)$ by the pair correlation $\mcC_{\alpha\beta}(\xx)$ (see Eq.~(\ref{eq:def_correlation})), which removes a term proportional to $\mcO(0)$.
In the last term of Eq.~(\ref{eq:current_correlations}), it is done by substracting $\delta_{\alpha\beta}\delta(\xx)C_{\beta\gamma}(\xx')$ to $C^{(3)}_{\alpha\beta\gamma}(\xx,\xx')$.
However, as we will compute the correlations in the Debye-Hückel limit where the odd correlations vanish, it is not necessary to remove this term.
Finally, note that replacing the correlation $C_{\alpha\beta}(\xx)$ by the pair correlation $\mcC_{\alpha\beta}(\xx)$ does not affect the electrostatic correction, allowing us to use the pair correlation in both corrections:
\begin{multline}\label{eq:current_correlations_pair}
\JJ = \sigma_0 \EE - \sum_{\alpha,\beta} qz_\alpha\kappa_\alpha \int\nabla V_{\alpha\beta}(\xx)\mcC_{\alpha\beta}(\xx)\dd\xx
+\sum_{\alpha,\beta} q^2z_\alpha z_\beta \int \mcO(\xx)\mcC_{\alpha\beta}(\xx)\dd\xx \,\EE 
\\-\sum_{\alpha,\beta,\gamma}qz_\alpha \int \mcO(\xx)\nabla V_{\beta\gamma}(\xx')C^{(3)}_{\alpha\beta\gamma}(\xx,\xx')\dd\xx\dd\xx'.
\end{multline}

Using the Parseval-Plancherel theorem and writing explicitly the time dependencies, we get
\begin{multline}
	\JJ(t)= \sigma_0 \boldsymbol{E}(t)+ \sum_{\alpha,\beta} q z_\alpha \kappa_\alpha \int i\kk \tilde{V}_{\alpha\beta}(\kk)\tilde{\mathcal{C}}_{\alpha\beta}(\kk,t)  \frac{\dd\kk}{(2\pi)^d} 
	\\+ \sum_{\alpha,\beta} q^2 z_\alpha z_\beta  \int \tilde{\mathcal{O}}(\kk) \tilde\mcC_{\alpha\beta}(\kk,t) \frac{\dd\kk}{(2\pi)^d} \boldsymbol{E}(t) 
	\\+ \sum_{\alpha,\beta,\gamma} q z_\alpha \int \tilde{\mathcal{O}}(\kk) i \kk' \tilde{V}_{\beta\gamma} (\kk') \tilde{C}^{(3)}_{\alpha\beta\gamma}(\kk,\kk',t) \frac{\dd\kk\dd\kk'}{(2\pi)^{2d}}.
	\label{eq:avg_J_k}
\end{multline}
We have used the fact that the Fourier transforms $\tilde V_{\alpha\beta}(\kk)=\frac{q^2z_\alpha z_\beta}{\epsilon k^2}$ and $\tilde O_{ij}(\kk)=\frac{1}{\eta k^2}(\delta_{ij}-\frac{k_ik_j}{k^2})$  are even: $\tilde V_{\alpha\beta}(\kk)=\tilde V_{\alpha\beta}(-\kk)$ and $\tilde \mcO(\kk)=\tilde\mcO(-\kk)$.

Now that we have expressed the correction to the bare current as a function of the correlations, we need to evaluate the correlations.

\subsection{Correlations in the Debye-Hückel limit}

The density correlations cannot be computed exactly.
To evaluate them, we assume small density fluctuations $|n_\alpha|\ll\bar\rho_\alpha$ and take the Debye-Hückel limit, which amounts to linearize the deterministic terms in the current Eq.~(\ref{eq:current_expanded}) and remove the fluctuations in front of the noise terms~\cite{Demery2016Conductivity,Peraud2017}.
Linearizing Eq.~(\ref{eq:current_expanded}) and plugging it into Eq.~(\ref{eq:continuity}), we get
\begin{equation}
\dot{n}_{\alpha}= \kappa_{\alpha}T\nabla^2 n_{\alpha} -\kappa_{\alpha}q z_\alpha \boldsymbol{E}\cdot \nabla n_\alpha   
+\kappa_{\alpha}\bar\rho_\alpha \nabla^2\left[\sum_\beta V_{\alpha\beta}*n_\beta\right]  +\sqrt{\kappa_{\alpha}T\bar\rho_{\alpha}} \nabla\cdot\zzeta_{\alpha}.
\end{equation}
Note that at this order, the terms coming from the hydrodynamic interaction disappear as the Oseen tensor and noise correlation function are divergence free.
The fluctuations $n_\alpha(\xx,t)$ are now Gaussian fields, so that odd correlations such as $C^{(3)}$ are zero.

We now write the dynamics in Fourier space:
\begin{equation}\label{eq:dyn_lin_fourier}
\dot{\tilde n}_{\alpha}= -\kappa_{\alpha}T k^2 \tilde n_{\alpha} + i \kappa_{\alpha}qz_\alpha \boldsymbol{E}\cdot \boldsymbol{k} \tilde n_\alpha   
-\kappa_{\alpha}\bar\rho_\alpha k^2 \sum_\beta \tilde V_{\alpha\beta} \tilde n_\beta    +
\sqrt{\kappa_{\alpha}T\bar\rho_{\alpha}} i\kk\cdot\tilde\zzeta_{\alpha}.
\end{equation}
The dynamics of the fluctuations, Eq.~(\ref{eq:dyn_lin_fourier}), can be written in a vectorial form:
\begin{equation}
\dot{\tilde n} = - R  A \tilde n + \chi,
\label{eq:vectorial}
\end{equation}
where $R_{\alpha\beta}(\kk)=\delta_{\alpha\beta}  \bar\rho_\alpha \kappa_\alpha k^2$ is the mobility matrix and $A_{\alpha\beta}(\kk) = \delta_{\alpha\beta} \frac{T}{\bar\rho_\alpha}\left(1+i\frac{z_\alpha q \boldsymbol{E}\cdot \boldsymbol{k}}{Tk^2}\right) + \tilde V_{\alpha\beta}$; we have introduced the scalar Gaussian noise $\chi_\alpha(\xx,t)$ with correlation 
\begin{equation}
	\langle\chi_\alpha(\kk,t)\chi_\beta(\kk',t')\rangle = 2 (2\pi)^d T R_{\alpha\beta}(\kk) \delta(\kk+\kk')\delta(t-t').
\end{equation}

In Fourier space, the correlation is given by
\begin{equation}
\langle \tilde n_\alpha(\kk,t) \tilde n_\beta(\kk',t)\rangle= (2\pi)^d \delta(\kk+\kk')\tilde C_{\alpha\beta} (\kk,t).
\end{equation}
Using the Itô product rule on Eq.~(\ref{eq:vectorial}) we find that the correlation $\tilde C$ follows (see Ref.~\cite{Gardiner2009}, Sec. 4.4)
\begin{equation}
	\dot{\tilde C}
	= 2TR -RA\tilde C- \tilde C A^* R,
	\label{eq:correlation_1}
\end{equation}
where $A^*$ is the the complex conjugate of $A$. 
This is a differential Lyapunov equation \cite{Behr2019Solution} and can be casted into a system of ODEs.

When the electric field is constant over the time interval $[0,t]$, which is the case for a switch on or a switch off of the field at $t=0$, the solution to Eq.~(\ref{eq:correlation_1}) is given by
\begin{equation}
\text{vec}\left(\tilde C(t)\right) = e^{-Mt}\left[\text{vec}\left(\tilde{C}(0)\right)-2TM^{-1}\text{vec}(R)\right] +2TM^{-1}\text{vec}(R),
\label{eq:systemODEsolution}
\end{equation}
where $M = \left[\mathbb{I}\otimes (RA)+(RA^*) \otimes  \mathbb{I}\right]$. 
The symbol $\otimes$ is the tensor product and $\text{vec}(\cdot)$ is the vectorization operator.
Another option would have been to integrate the differential linear equation (\ref{eq:vectorial}) and then take the average, leading to the solution (\ref{eq:systemODEsolution})~\cite{Dean2012,Dean2014,Mahdisoltani2021Transient}.

\subsection{Binary monovalent electrolytes: dimensionless form}

We restric ourselves to the case of a binary monovalent electrolyte, where both species have the same mobility: $\alpha=\{+,-\}$, $z_+=-z_-=1$, $\bar\rho_\alpha=\bar\rho$, and $\kappa_\alpha=\kappa$.

We nondimensionalize Eq.~(\ref{eq:correlation_1}) by setting $\tilde{C}= \bar \rho \tilde c$ and $\kk=\ss/\lambda\ind{D}$ where $\ldeb=\sqrt{T \epsilon/(2q^2 \bar{\rho})}$ is the Debye length. 
Then we rescale time by the Debye time $t\ind{D} = \ldeb^2 /(\kappa T)$, $t=t\ind{D}\tau$. 
We rewrite the external field to separate the magnitude from the time dependence $E(t) = E_0 g(t)$ and introduce the dimensionless field $f= q \ldeb E_0/T$.
The rescaled correlation $c_{\alpha\beta}(\tau)$ follows
\begin{equation}\label{eq:evol_cor_adim}
\dot {\tilde c} = 2s^2-\omega \tilde c-\tilde c\omega^*,
\end{equation}
where we have introduced the matrix $\omega$, which is a dimensionless version of $RA$:
\begin{equation}
\omega_{\alpha\beta}(\ss) = \delta_{\alpha\beta}\left(s^2+iz_\alpha fs_x\right)+\frac{z_\alpha z_\beta}{2}.
\end{equation}
Explicit expressions of the dimensionless correlations from Eq.~(\ref{eq:systemODEsolution}) are given in App.~\ref{app:correlations} for the NESS to equilibrium and equilibrium to NESS cases.

 
Applying the same scaling to the current $J(\tau)$ (Eq.~(\ref{eq:avg_J_k})) we find:
\begin{equation}
	\frac{J(\tau)}{\sigma_0 E_0} =g(\tau)+\frac{ 1 }{\bar{\rho} \ldeb^3  } \gamma\ind{el}(\tau,f)+g(t)\frac{r_s}{\ldeb} \gamma\ind{hyd}(\tau,f),
	\label{eq:corrections}
\end{equation}
where $r_s=(6\pi\eta\kappa)^{-1}$ is the hydrodynamic radius of the charged particles, and the electrostatic and hydrodynamic corrections read, respectively,
\begin{align}
\gamma\ind{el}(\tau,f) &= - \frac{1}{16 \pi^2 f} \int_{0}^{\infty}  \dd s\int_{-1}^{1} \dd u\  i s u \sum_{\alpha,\beta}z_\alpha \left[\tilde c_{\alpha\beta}(s,u,\tau,f)-\delta_{\alpha\beta}\right], \label{eq:gamma_el}\\
\gamma\ind{hyd}(\tau,f) &= \frac{3}{4 \pi } \int_{0}^{\infty}  \dd s\int_{-1}^{1} \dd u  \left( 1- u^2 \right)\sum_{\alpha,\beta}z_\alpha z_\beta \left[\tilde c_{\alpha\beta}(s,u,\tau,f)-\delta_{\alpha\beta}\right]. \label{eq:gamma_hyd}
\end{align}
We have introduced the variable $u=s_x/s$.
The dimensionless parameters in front of the correction terms in Eq.~(\ref{eq:corrections}) imply that none can by neglected, and we study them separately.
Note that the electrostatic correction involves the odd part of the correlations, as the prefactor is odd in the variable $u$, while the hydrodynamic correction involves the even part of the correlation.
Note also that the hydrodynamic correction is multiplied by the time dependence of the electric field, $g(t)$, hence it is absent in the transition from NESS to equilibrium. 

The corrections in the steady state have been computed previously~\cite{Onsager1957Wien,Demery2016Conductivity}:
\begin{align}
\gamma\ind{el}^\infty &= - \frac{1}{32 \pi f^3}\left[f \sqrt{f^2+1}-\sqrt{2} f+\tan ^{-1}\left(\sqrt{2} f\right)-\tan ^{-1}\left(\frac{f}{\sqrt{f^2+1}}\right)\right], \label{eq:gamma_el_inf}\\
\gamma\ind{hyd}^\infty &=-\frac{1}{\sqrt{2}} -\frac{3 \left(\sqrt{2}-\sqrt{f^2+1}\right)}{8 f^2}-\frac{3}{4 f} \sinh ^{-1}(f)\nonumber
	\\&\qquad+\frac{3(1+2f^2)}{8 f^3} \left[\tan ^{-1}\left(\sqrt{2} f\right)-\tan ^{-1}\left(\frac{f}{\sqrt{f^2+1}}\right)\right].\label{eq:gamma_hyd_inf}
\end{align}

\subsection{Correlations from equilibrium to NESS and back}\label{}

\begin{figure}
	\centering
	\includegraphics[width=\linewidth]{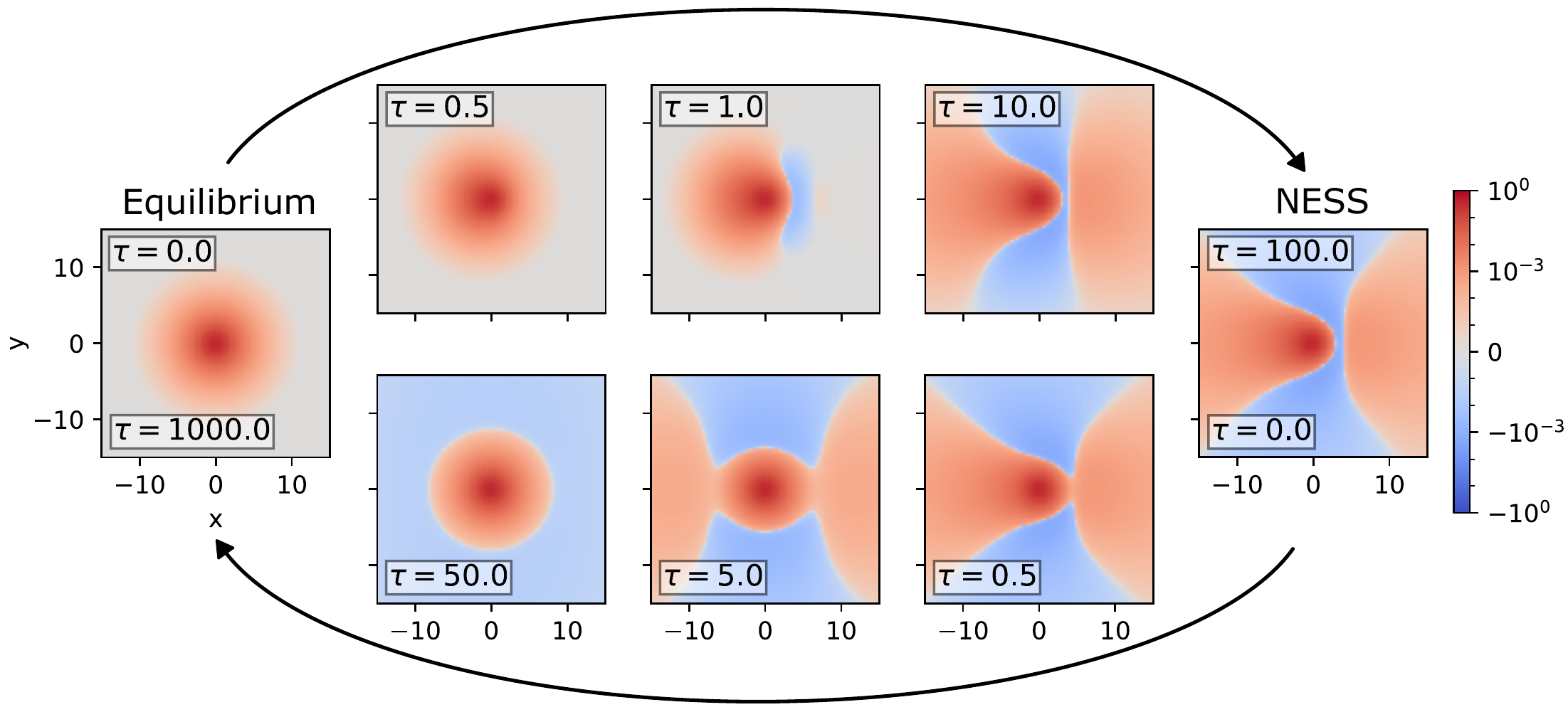}
	\caption{Evolution of the correlation $c_{+-}(\xx)$ in a binary monovalent electrolyte for $f=1$, from equilibrium to NESS (top row), and from NESS to equilibrium (bottom row). The field is oriented along $x$.}
	\label{fig:nesstoeq}
\end{figure}

We present the evolution of the pair correlations for different species, $c_{+-}$, for the transitions from equilibrium to NESS, and from NESS to equilibrium, in Fig.~\ref{fig:nesstoeq}.
We see that the trajectory from NESS to equilibrium is not the inverse of the trajectory from equilibrium to NESS.
In particular, it seems that from NESS to equilibrium, the correlation quickly becomes symmetric before slowly relaxing to its equilibrium value.
In the next sections, we focus on the behavior of the conductivity, which we finally compare to the evolution of the correlation.

\section{From NESS to equilibrium}
\label{sec:ness_eq}

We start by studying the dynamics of the conductivity as a response to a sudden switch off of the electric field (NESS to equilibrium).
We solve Eq.~(\ref{eq:evol_cor_adim}) under $f=0$ with the initial condition being the steady state solution to Eq.~(\ref{eq:evol_cor_adim})
for a finite value of $f$.
After the switch off of the field, the bare current is zero, hence the only current comes from the out of equilibrium correlations through the electrostatic correction, as there is no hydrodynamic correction in this case.

\begin{figure}
\begin{center}
	\includegraphics{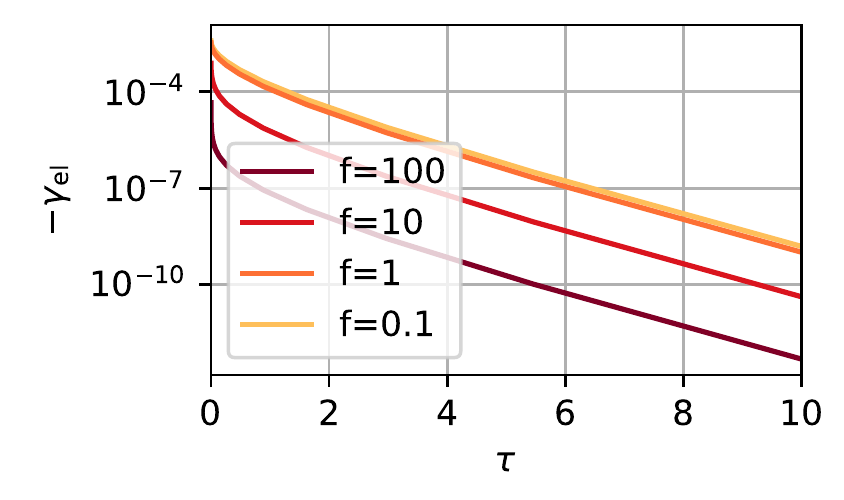}
	\caption{Electrostatic correction term $\gamma\ind{el}(\tau)$ from NESS to equilibrium (Eq.~(\ref{eq:nte_el}), integrated numerically) as a function of time for different values of the field $f$.}
	\label{fig:equilibrium_time}
\end{center}
\end{figure}

By plugging the solution of Eq.~(\ref{eq:evol_cor_adim}) as described to Eq.~(\ref{eq:gamma_el}), we find that the electrostatic correction is 
\begin{equation}
	\gamma\ind{el}(\tau)=-\frac{1}{8 \pi ^2 f }\int_{0}^{\infty} \dd s\int_{-1}^{1}\dd u\frac{s^2 u^2 e^{-\left(2 s^2+1\right) \tau}}{ \left(2 s^2+1\right) \left(f^2 u^2+s^2+1\right)}.
	\label{eq:nte_el}
\end{equation}
This integral is integrated numerically with the package Quadpack implemented in SciPy~\cite{Piessens2012Quadpack,Virtanen2020Scipy}; it is plotted as a function of time in Fig.~\ref{fig:equilibrium_time}.
Differentiating Eq.~(\ref{eq:nte_el}), one can show that the correction decays monotonically, as seen in Fig.~\ref{fig:equilibrium_time}.

\begin{figure}
\begin{center}
	\includegraphics{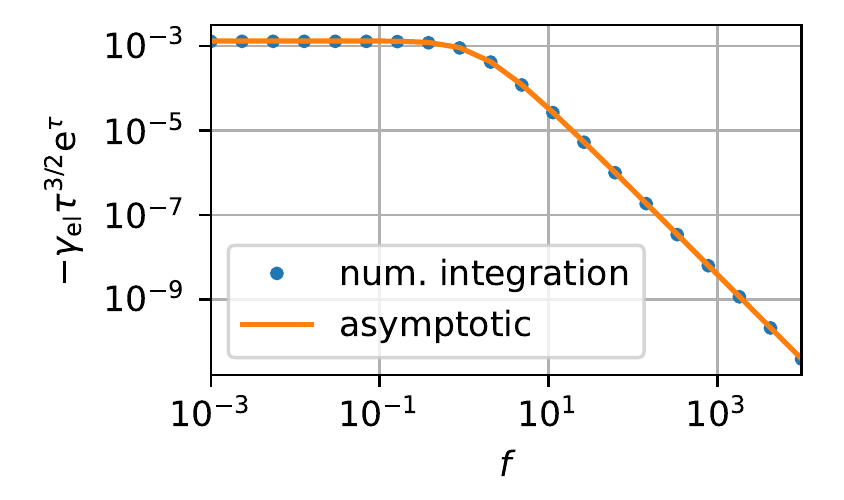}
	\caption{Field dependency of the electrostatic correction $\gamma\ind{el}(\tau)$ at large time. Analytical result (Eq.~(\ref{eq:nte_el_lt}), solid line) and from numerical integration of Eq.~(\ref{eq:nte_el}) at $\tau=100$ (dots).}
	\label{fig:nte_el_ltpref}
\end{center}
\end{figure}

At short time, the electrostatic correction behaves as
\begin{equation}
\gamma\ind{el}(\tau) \underset{\tau \to 0}{\sim}\frac{\sqrt{\tau}}{12 \sqrt{2} \pi ^{3/2}}.
\end{equation}
At long times, it decays faster than exponentially,
\begin{equation}
\gamma\ind{el}(\tau)\underset{\tau\to\infty}{\sim}
		-\frac{e^{-\tau} \left[f-\tan^{-1}(f)\right]}{32 \sqrt{2} \pi ^{3/2}  f^3 \tau^{3/2}},\label{eq:nte_el_lt}
\end{equation}
with a field dependent prefactor that is constant at weak field and decays as $f^{-2}$ at large field (Fig.~\ref{fig:nte_el_ltpref}).

As a conclusion, after a switch off of the external field there is a recoil coming from the electrostatic correction.
This recoil decays exponentially on the time scale of the Debye time.
This decay is compatible with the evolution of the correlations (Fig.~\ref{fig:nesstoeq}): the antisymmetric part, which enters in the electrostatic  correction, indeed seems to decay on the time scale of the Debye time.

\section{From equilibrium to NESS}
\label{sec:eq_ness}

We now turn to the dynamics of the conductivity after a sudden switch on of the electric field, the system being initially in its equilibrium state.
Here both electrostatic and hydrodynamic corrections are present, and we study them separately.

\subsection{Electrostatic correction}

The integrand of $\gamma\ind{el}$ in Eq.~(\ref{eq:gamma_el}), which we denote $y(s,u,f,\tau)$ is obtained from the solution (\ref{eq:systemODEsolution}):
\begin{multline}
\label{eq:integrand_etn_el}
y(s,u,f,\tau) =
\frac{2f s^2 u^2 e^{-\tau \left(2s^2+1\right) }}{\left(s^2+1\right) \left(2s^2+1\right) \left(f^2 u^2+s^2+1\right) \left(1-4f^2 s^2 u^2\right)}  
\bigg[- \left(f^2 u^2+s^2+1\right) \\
+f^2u^2  \left(2s^2+1\right)\left(\left(2s^2+1\right) \cosh\left( \tau \sqrt{1-4f^2 s^2 u^2}\right) +\sqrt{1-4f^2 s^2 u^2} \sinh\left( \tau \sqrt{1-4f^2 s^2 u^2}\right)\right)\bigg]\\
+\frac{2f s^2 u^2 }{ \left(2s^2+1\right) \left(f^2 u^2+s^2+1\right) }.
\end{multline}
The last term in $y$ is independent of time and corresponds, after integration, to the steady state result $\gamma\ind{el}^\infty$ (Eq.~(\ref{eq:gamma_el_inf})).
Note that the integrand is regular at the pole corresponding to $4s^2u^2f^2=1$ (App.~\ref{appendix:limit}).

\begin{figure}
	\begin{center} 
	\includegraphics[width=\linewidth]{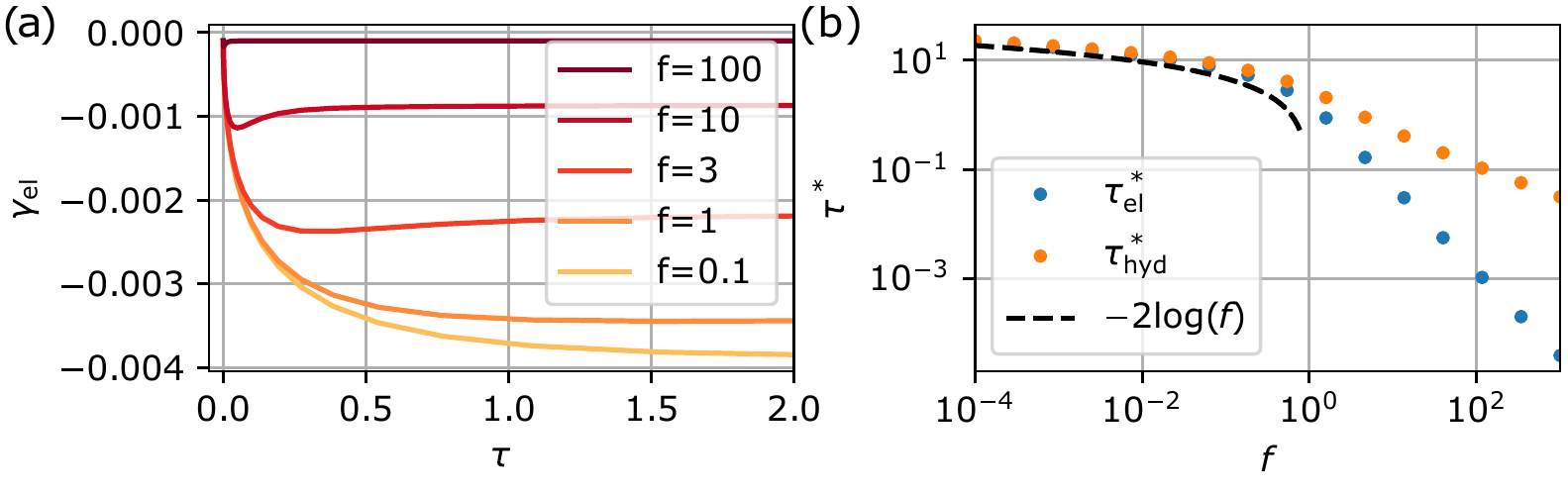} 
	\caption{(a) Electrostatic correction $\gamma\ind{el}(\tau)$ from equilibrium to NESS from numerical integration of Eq.~(\ref{eq:gamma_el}), for different values of the field $f$.
	         (b) Times $\tau\ind{el}^*(f)$ and $\tau\ind{hyd}^*(f)$ of the extrema of the electrostatic and hydrodynamic corrections, as a function of the magnitude of the external field $f$. Numerical evaluation of the time location of the extrema for the electrostatic correction (blue points) and for the hydrodynamic correction (orange points), and weak field asymptotics (\ref{eq:tstar}) (solid line).
	\label{fig:gamma}}
	\end{center}
\end{figure}

The electrostatic correction is integrated numerically and shown as a function of time in Fig.~\ref{fig:gamma}(a) for different values of the field.
Contrary to the monotonic behavior from NESS to equilibrium, here we observe an overshoot of the conductivity, which has a global minimum, lower than the steady state value $\gamma\ind{el}^\infty$.
The time location of the minimum as a function of the field, $\tau\ind{el}^*(f)$, is shown in Fig.~\ref{fig:gamma}(b): we see that the minimum is always present and that it occurs sooner and sooner as the field increases.
We also observe in Fig.~\ref{fig:gamma}(a) that the correction converges to a well defined limit at weak field, which corresponds to the linear response of the system.
We now focus on the short and long time behaviors of the correction.

\subsubsection{Short time limit}

\begin{figure}
\begin{center} 
\includegraphics[width=0.5\linewidth]{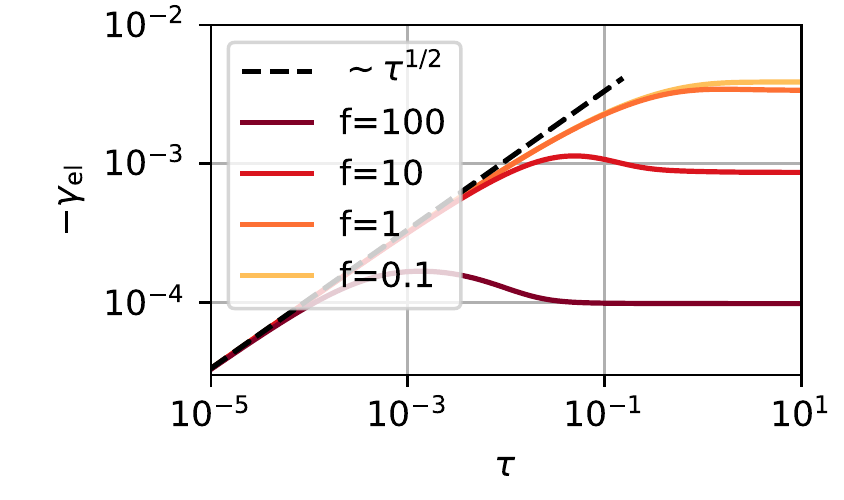}
\caption{Short time behavior of the electrostatic correction. Numerical integration for different values of the field (solid lines) and short time asymptotics (Eq.~(\ref{eq:gamma_short}), dashed line).}
\label{fig:gamma_short}
\end{center}
\end{figure}

The behavior of the conductivity can obtained by the change of variables $w=\sqrt{\tau}s$ followed by a Taylor expansion of the integrand to lowest order around $\tau\to 0$, leading to
\begin{equation}
\gamma\ind{el}(\tau) \underset{\tau \to 0}{\sim}  - \frac{1}{16 \pi^2} \sqrt{\tau} \int_{0}^{\infty}\dd w \int_{-1}^{1}\dd u\,  u^2\frac{  1-e^{-  2w^{2} }}{w^{2}}.
\end{equation}
Note that to obtain the asymptotic form, the steady state part of the integrand in Eq.~(\ref{eq:gamma_el}) cannot be computed separately, but has to be part of the expanded expression.
Evaluating the integral gives
\begin{equation}
\gamma\ind{el}(\tau)  \underset{\tau \to 0}{\sim}  - \frac{\sqrt{\tau}}{12\sqrt{2}\pi^{\frac{3}{2}}}.
\label{eq:gamma_short}
\end{equation}
The short time asymptotics is compared to the numerical integration in Fig.~\ref{fig:gamma_short}.
We recover the square root dependence observed from NESS to equilibrium.
Interestingly, the short time asymptotics does not depend on the field; as this is the correction to the conductivity, it means that the short time response is linear.
However, the higher the field, the sooner the conductivity departs from the short time asymptotics.

\subsubsection{Long time limit}

\begin{figure}
	\begin{center} 
	\includegraphics[width=\linewidth]{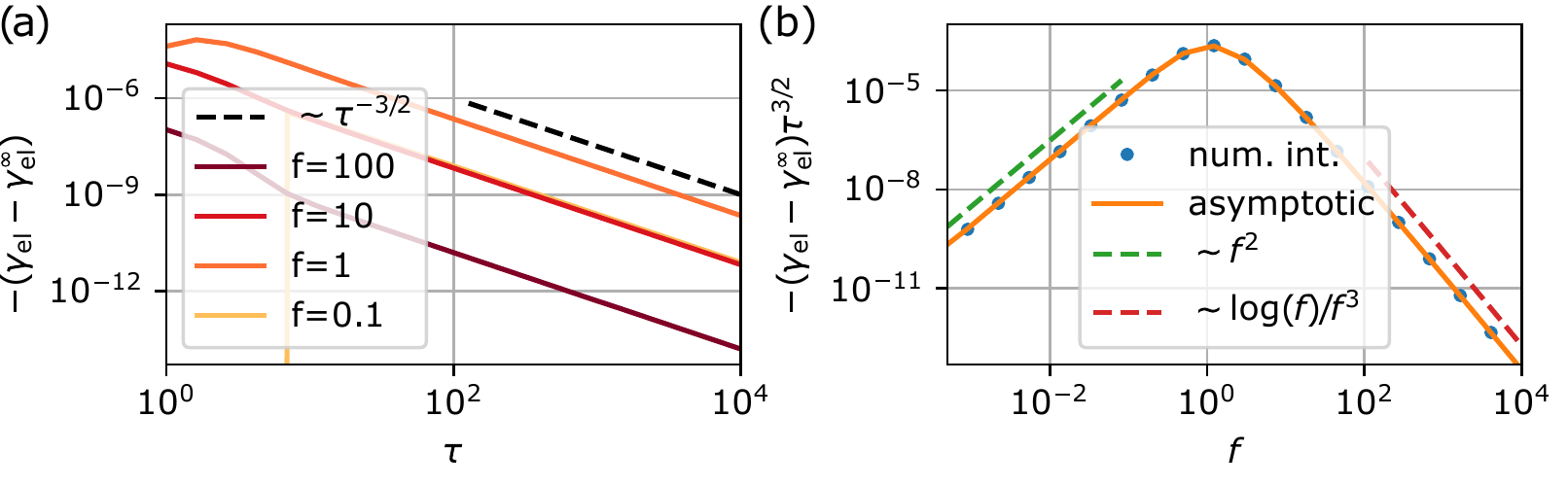}
	\caption{Long time behavior of the electrostatic correction.
	         (a) Difference with the stationary value as a function of time for different values of the field (solid lines) and $\tau^{-3/2}$ power law (dashed line). 
	         Note that the correction is regular, the apparent jump for $f=0.1$ is due to the logarithmic scale.
	         (b) Prefactor of the algebraic decay as a function of the field $f$ from numerical integration (blue points), from the long time asymptotics (Eq.~(\ref{eq:gamma_long}), solid line), and small and large field asymptotics (dashed lines).} 
	\label{fig:gamma_long}
	\end{center}
\end{figure}

To estimate the behavior of the conductivity at long times, we first separate the time independent part in the integrand (\ref{eq:integrand_etn_el}), which integrates to $\gamma\ind{el}^\infty$.
The remaining time dependent part of the integrand gives us access to the large time asymptotic behavior of the conductivity.
We perform the same change of variables $w=\sqrt{\tau}s$ and then expand the integrand to the lowest order around $\tau\to \infty$, yielding
\begin{equation}
\gamma\ind{el}(\tau)-\gamma\ind{el}^{\infty} \underset{\tau \to \infty}{\sim}  -  \frac{1}{16 \pi^2}\frac{1}{\tau^{3/2}}  \int_{0}^{\infty}\dd w \int_{-1}^{1}\dd u \frac{2 w^2 u^4 f^4 e^{-2w^2} }{  \left(f^2 u^2+1\right) }
 e^{  -2 f^2 w^2 u^2}.
\end{equation}
Evaluating the integral gives
\begin{equation}
\label{eq:gamma_long}
\gamma\ind{el}(\tau) -\gamma\ind{el}^{\infty}\underset{\tau \to \infty}{\sim}  - \frac{3 \left(f^2+1\right)^{3/2} \sinh ^{-1}(f)-4 f^3-3 f}{96 \sqrt{2} \pi ^{3/2}   f^3 \left(f^2+1\right)^{3/2} \tau^{3/2}}.
\end{equation}
The algebraic decay of the electrostatic correction is visible in the numerical evaluation in Fig.~\ref{fig:gamma_long}(a); the prefactor obtained from the numerical evaluation is compared to the expression (\ref{eq:gamma_long}) in Fig.~\ref{fig:gamma_long}(b).

Our main observation is that the relaxation of the electrostatic correction towards its stationary value is algebraic, $\gamma\ind{el}(\tau) -\gamma\ind{el}^{\infty}\sim\tau^{-3/2}$, contrary to the exponential relaxation when going from NESS to equilibrium.
This algebraic behavior is reminiscent of the one seen for the relaxation of the long range force between two boundaries of an electrolyte in the same configuration~\cite{Mahdisoltani2021Transient}.
As the prefactor of the algebraic decay goes to zero as the field goes to zero, it is a non-linear effect.
To better understand the non-linear effects, we now focus on the weak field limit.

\subsubsection{Weak field limit}

The weak field limit of the electrostatic correction can be obtained by expanding the integrand (\ref{eq:integrand_etn_el}), leading to
\begin{equation}
\gamma\ind{el} = \sum_{n=0}^\infty f^{2n}\gamma\ind{el}^{(2n)}.
\end{equation}
The first two terms are
\begin{align}
\gamma\ind{el}^{(0)}(\tau) &= -\frac{  \sqrt{2} \erfc{\left(\sqrt{\tau} \right)} - 2 e^{\tau} \erfc{\left(\sqrt{2\tau} \right)} +2- \sqrt{2}}{48 \pi},\label{eq:gamma_etn_el_0} \\
\gamma\ind{el}^{(2)}(\tau) &=\frac{1}{20 \pi}
	\left[\frac{3-2 \sqrt{2}}{4} +\frac{4 \tau-8 e^\tau+2 e^{2\tau}+3}{4}  \erfc\left(\sqrt{2\tau}\right)\right.\nonumber\\
	&\qquad\left.+\frac{\sqrt{\pi } \erfc\left(\sqrt{\tau}\right)-e^{-2 \tau} \sqrt{\tau}}{\sqrt{2 \pi }} 
	-\frac{\left(1-e^{-\tau}\right)^2}{2 \sqrt{2 \pi \tau } }
	\right].
\end{align}

The long time behavior of the lowest order term (Eq.~(\ref{eq:gamma_etn_el_0})) is
\begin{equation}\label{eq:gamma_etn_el_0_lt}
\gamma\ind{el}^{(0)}(\tau)-\gamma\ind{el}^{\infty(0)} \underset{\tau \to \infty}{\sim} \frac{e^{-\tau}}{96\sqrt{2}\pi^{3/2}\tau^{3/2}}.
\end{equation}
It decays exponentially, confirming that the algebraic decay at finite field is rooted in the non-linear response.
Moreover, the asymptotics (\ref{eq:gamma_etn_el_0_lt}) matches exactly the long time asymptotics when going from NESS to equilibrium (obtained from a weak field expansion of Eq.~(\ref{eq:nte_el_lt})).
Indeed, expanding the equation (\ref{eq:evol_cor_adim}) for the correlation at weak field reveals that the NESS to equilibrium and equilibrium to NESS trajectories are identical in the linear regime.

At the next order, the algebraic behavior is recovered,
\begin{equation}
\label{eq:gamma_etn_el_2_lt}
\gamma\ind{el}^{(2)}(\tau)-\gamma\ind{el}^{\infty(2)} \underset{\tau \to \infty}{\sim} -\frac{1}{160\sqrt{2}\pi^{3/2}\tau^{3/2}},
\end{equation}
in agreement with the $f^2$ dependence of the prefactor of the algebraic decay (Fig.~\ref{fig:gamma_long}(b)). 

The lowest order term, $\gamma\ind{el}^{(0)}(\tau)$, is a decreasing function of $\tau$, while the next order, $\gamma\ind{el}^{(2)}(\tau)$, is increasing.
The time location of the minimum, $\tau\ind{el}^*$, can be obtained at low field by comparing these two terms. 
As $\tau\ind{el}^*$ seems to diverge as the field goes to zero (Fig.~\ref{fig:gamma}(b)), it is sufficient to use the long time asymptotics, Eqs.~(\ref{eq:gamma_etn_el_0_lt}, \ref{eq:gamma_etn_el_2_lt}).
Differentiating the long time asymptotics of $\gamma\ind{el}^{(0)}(\tau)+f^2\gamma\ind{el}^{(2)}(\tau)$ gives the following equation for the time location of the minimum: $9f^2 e^{\tau}-10\tau-15=0$.
To leading order in $f$, the solution is
\begin{equation}\label{eq:tstar}
\tau\ind{el}^*(f)\underset{f \to 0}{\sim}-2\log(f).
\end{equation}
This asymptotic behavior is compared to the numerical evaluation of the time location of the minimum in Fig.~\ref{fig:gamma}(b).

\subsection{Hydrodynamic correction}

The integrand of $\gamma\ind{hyd}$ in Eq.~(\ref{eq:gamma_hyd}), which we denote $h(s,u,f,\tau)$ is obtained from the solution (\ref{eq:systemODEsolution}):
\begin{multline}
		h(s,u,f,\tau)=-\frac{f^2 s^2 u^2 \left(u^2-1\right) e^{-\tau \left(2 s^2+1\right)}}{\left(s^2+1\right) \left(2 s^2+1\right) \left(f^2 u^2+s^2+1\right) \left(1-4 f^2 s^2 u^2\right)^{3/2}}
	\\\times\bigg(-4 \left(f^2 u^2+s^2+1\right) \sqrt{1-4 f^2 s^2 u^2} 
	\\+\left(2 s^2+1\right) \left[2 f^2 u^2 \left(\sqrt{1-4 f^2 s^2 u^2}+2 s^2\right)+\sqrt{1-4 f^2 s^2 u^2}-1\right] e^{ \tau \sqrt{1-4 f^2 s^2 u^2}}
	\\+\left(2 s^2+1\right) \left[2 f^2 u^2 \left(\sqrt{1-4 f^2 s^2 u^2}-2 s^2\right)+\sqrt{1-4 f^2 s^2 u^2}+1\right]e^{-\tau\sqrt{1-4 f^2 s^2 u^2}}\bigg)
	\\+\frac{2 \left(u^2-1\right) \left(f^2 u^2+2 s^2+1\right)}{\left(2 s^2+1\right) \left(f^2 u^2+s^2+1\right)}.
	\label{eq:h}
\end{multline}
The last term in $h$ is independent of time and corresponds to the steady state correction $\gamma\ind{hyd}^\infty$ (Eq.~(\ref{eq:gamma_hyd_inf})).

\begin{figure}
\begin{center} 
\includegraphics[width=0.5\linewidth]{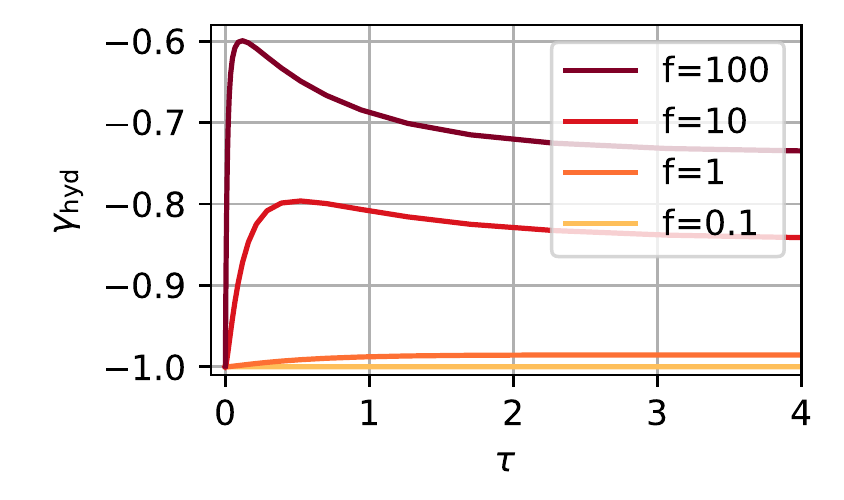}
\caption{Hydrodynamic correction $\gamma\ind{hyd}(\tau)$ from equilibrium to NESS from numerical integration of Eq.~(\ref{eq:gamma_hyd}) for different values of the field $f$.} 
\label{fig:gamma_hyd}
\end{center}
\end{figure}

The hydrodynamic correction is integrated numerically and shown as a function of time in Fig.~\ref{fig:gamma_hyd}.
The main difference with the electrostatic correction is the finite value at $\tau=0$. 
This is due to the fact that the hydrodynamic correction involves the even part of the correlation, which is finite for the equilibrium initial condition.
As for the electrostatic correction, the relaxation towards the stationary value is non-monotonic; the time location of the maximum as a function of the field is plotted in Fig.~\ref{fig:gamma}(b).
We now study the short and long time behaviors of the correction.

\subsubsection{Short time limit}

\begin{figure}
\begin{center} 
\includegraphics[width=0.5\linewidth]{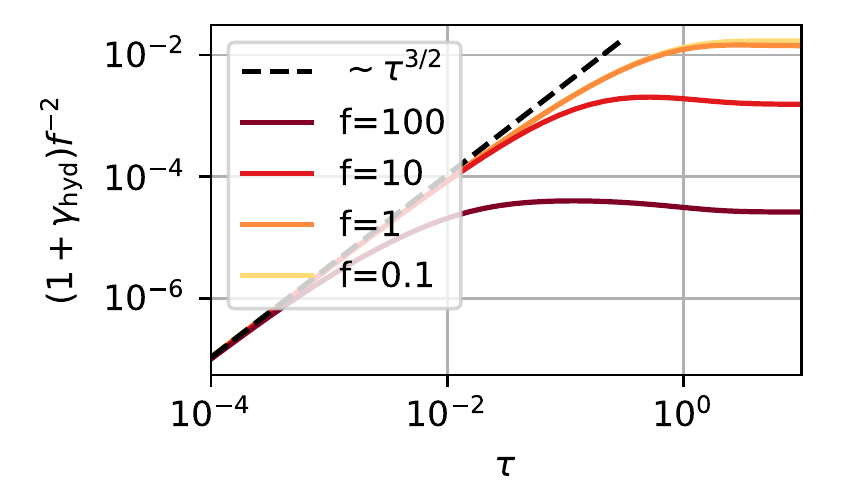}
\caption{Short time behavior of the hydrodynamic correction. Numerical integration for different values of the field (solid lines) and short time asymptotics (Eq.~(\ref{eq:gamma_hyd_short_t}), dashed line).
}
\label{fig:etn_hyd_short}
\end{center}
\end{figure}

Using the same method as for the electrostatic correction, we find the short time behavior of the hydrodynamic correction.
We find $\gamma\ind{hyd}(\tau)\underset{\tau\to 0}{\to}-1$ and
\begin{equation}
\gamma\ind{hyd}(\tau)+1 \underset{\tau \to 0}{\sim} \frac{2}{15} \sqrt{\frac{2}{\pi }} f^2 \tau^{3/2};
\label{eq:gamma_hyd_short_t}
\end{equation}
it is compared to the numerical integration in Fig.~\ref{fig:etn_hyd_short}.
The short time evolution is slower than for the electrostatic correction and depends on the field.
We also observe that the departure from the short time asymptotics occurs sooner for a larger field.

\subsubsection{Long time limit}

To estimate the behavior of the hydrodynamic correction at long times, we first separate the time independent part in the integrand (\ref{eq:h}), which integrates to $\gamma\ind{hyd}^\infty$.
The remaining time dependent part of the integrand gives us access to the large time asymptotic behavior of the conductivity.
We perform the same change of variables, $w=\sqrt{\tau}s$ and then expand the integrand to the lowest order around $\tau\to \infty$, yielding
\begin{equation}
\gamma\ind{hyd}(\tau)-\gamma\ind{hyd}^\infty \underset{\tau \to \infty}{\sim} \frac{1}{16 \sqrt{2 \pi } \tau^{3/2}}\left[ \frac{15+6f^2}{ f^3 }\sinh ^{-1}(f)-\frac{15+11f^2}{ f^2  \sqrt{f^2+1 }}\right].
\label{eq:gamma_hyd_long_t}
\end{equation}
This asymptotics is compared to the numerical integration in Fig.~\ref{fig:etn_hyd_long}(a).
We find again an algebraic relaxation towards the stationnary value, with the same exponent $3/2$ as for the electrostatic correction.
Similarly, the prefactor goes to zero as the field goes to zero (Fig.~\ref{fig:etn_hyd_long}(b)), indicating that the algebraic decay is a non-linear effect.

\subsubsection{Weak field limit}

The weak field limit of the hydrodynamic correction can be obtained by expanding the integrand (\ref{eq:h}), leading to
\begin{equation}
	\gamma\ind{hyd} = \sum_{n=0}^\infty f^{2n}\gamma\ind{hyd}^{(2n)}.
\end{equation}
The lowest order term is $\gamma\ind{hyd}^{(0)}(\tau)=-1$.
It relaxes instantaneously, so that its difference with its value in the stationary state is zero, mirroring the absence of the hydrodynamic correction when going from NESS to equilibrium.
This instantaneous relaxation also shows that the algebraic decay is, as for the electrostatic correction, a non-linear effect.

\begin{figure}
	\begin{center} 
		\includegraphics[width=\linewidth]{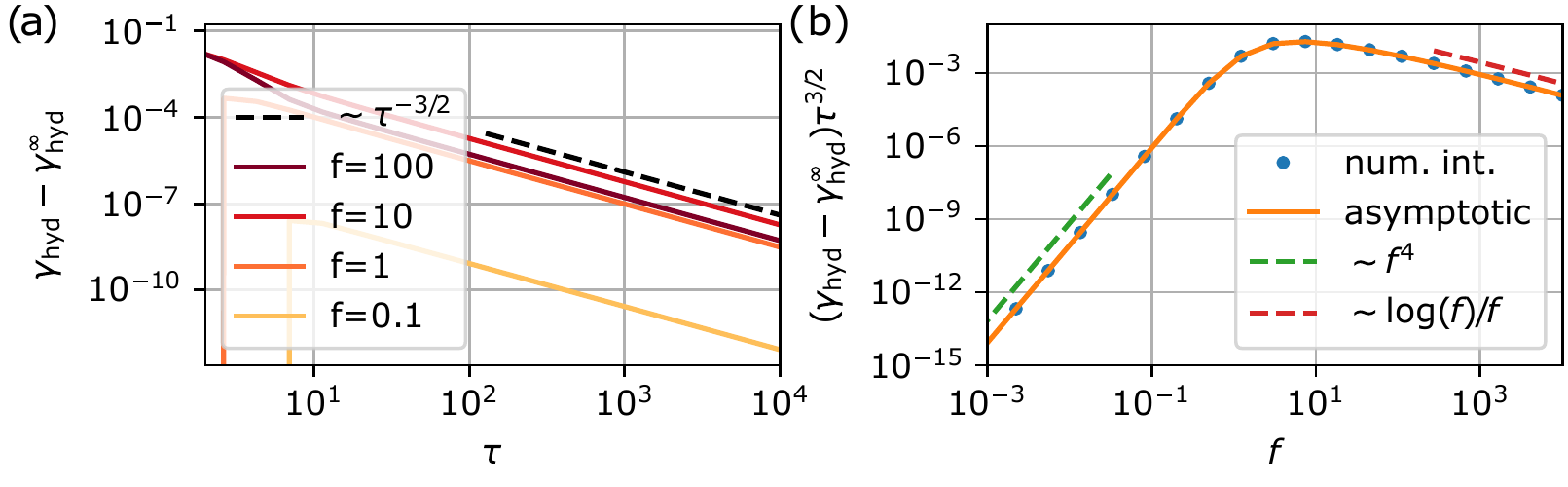}
		\caption{Long time behavior of the hydrodynamic correction.
		         (a) Difference with the stationary value as a function of time for different values of the field (solid lines) and $\tau^{-3/2}$ power law (dashed line).
		         (b) Prefactor of the algebraic decay as a function of the field from numerical integration (blue points), from the long time asymptotics (Eq.~(\ref{eq:gamma_hyd_long_t})), and small and large field asymptotics (dashed lines).}
    \label{fig:etn_hyd_long}
	\end{center}
\end{figure}

The following terms are
\begin{align}
	\gamma\ind{hyd}^{(2)} &=\frac{3-2\sqrt{2}}{10}+\frac{1}{10} \left[\left(4 \tau-4 e^\tau+1\right) \erfc\left(\sqrt{2\tau}\right)+2 \sqrt{2} \erfc\left(\sqrt{\tau}\right)-2 \sqrt{\frac{2}{\pi }} e^{-2 \tau} \sqrt{\tau}\right],\label{eq:gamma_etn_hyd_2}\\
	\gamma\ind{hyd}^{(4)} &=\frac{3}{280} \left(16 \sqrt{2}-23\right)+\frac{3}{280} \bigg[-\left(16 \tau ^2+48 \tau -64 e^{\tau }+8 e^{2 \tau }+33\right) \erfc\left(\sqrt{2\tau}\right) \nonumber\\
	& \qquad -16 \sqrt{2} \erfc\left(\sqrt{\tau }\right)+2 \sqrt{\frac{2}{\pi }} e^{-2 \tau }\frac{ 4 \tau ^2+11 \tau -8 e^{\tau }+2 e^{2 \tau }+6}{\sqrt{\tau }}\bigg].
\end{align}
They behave at long time as
\begin{align}
\gamma\ind{hyd}^{(2)}(\tau)-\gamma\ind{hyd}^{\infty(2)} &\underset{\tau \to \infty}{\sim} \frac{e^{-2 \tau } -2 e^{-\tau }}{20 \sqrt{2 \pi } \tau ^{3/2}},\label{eq:gamma_etn_hyd_2_lt}\\
\gamma\ind{hyd}^{(4)}(\tau)-\gamma\ind{hyd}^{\infty(4)} &\underset{\tau \to \infty}{\sim} \frac{3}{140 \sqrt{2 \pi } \tau ^{3/2}}.\label{eq:gamma_etn_hyd_4_lt}
\end{align}
Here, the algebraic decays appears at the fourth order in the field, in agreement with the $f^4$ dependence of the prefactor of the algebraic decay (Fig.~\ref{fig:etn_hyd_long}(b)). 

Finally, as for the electrostatic correction, comparing the orders 2 and 4 (Eqs.~(\ref{eq:gamma_etn_hyd_2_lt}, \ref{eq:gamma_etn_hyd_4_lt})) allows to find the asymptotic behavior of the time location of the minimum at weak field:
\begin{equation}\label{eq:tstar}
	\tau\ind{hyd}^*(f)\underset{f \to 0}{\sim}-2\log(f).
\end{equation}
This asymptotics is the same as for the electrostatic correction Fig.~\ref{fig:gamma}(b).

\section{Conclusion}
\label{sec:conclusion}

We have computed the transient ionic correlations in an electrolyte under a sudden switch on or switch off of an external electric field of arbitrary magnitude using linearized SDFT.
We have shown that the correlations do not follow the same trajectory when the field is switched on or switched off.
We have deduced the electrostatic and hydrodynamic corrections from the ionic correlations.
When the field is switched off, the hydrodynamic correction is absent and the electrostatic correction decays exponentially.
This fast decay seems to correspond to the fast decay of the odd part of the correlations; in contrast, the even part of the correlations, which is not involved in the electrostatic correction, decays slower.
On the contrary, when the field is switched on, we found that both corrections relax algebraically towards their stationnary value, with the same exponent.
In the linear response regime, an exponential relaxation is recovered when the field is switched on, showing that the algebraic relaxation is a non-linear effect.

The electrostatic and hydrodynamic corrections to the conductivity are given by the ionic correlations (Eqs.~(\ref{eq:gamma_el},\ref{eq:gamma_hyd})), and we have discussed the relation between the temporal behaviors of the correlations and the corrections.
However, our discussion of the evolution of the correlations has been limited to qualitative aspects, which is due to their poor characterization in the NESS.
Indeed, while the correlations are of the Yukawa form at equilibrium, isotropic and  exponentially decaying with distance, they are long ranged in the NESS~\cite{Mahdisoltani2021}.
The algebraic decay of the correlations out of equilibrium has been shown to give rise to long ranged forces, but it has not been characterized precisely in the stationnary state, nor in the transient regime. 
A characterization such as the one obtained for a driven binary mixture with short range interactions~\cite{Poncet2017} would help to pinpoint the role of the correlations in the asymmetric behavior unveiled here when the field is switched on or off.

Finally, it would be interesting to understand how the results obtained here for a sudden switch on or switch off of the electric field translate to different time dependencies, such as an AC driving~\cite{Richter2020}.

\section*{Acknowledgements}
We thank David S. Dean for stimulating discussions,
Peter C. W. Holdsworth for pointing to the reversibilty of the trajectory for  a weak field and Yael Avni and David Andelman for explaining why the hydrodynamic interactions do not enter in the linearized equations.


\paragraph{Funding information}
We acknowledge financial support from Ecole Doctorale ED564 ``Physique en Ile de France'' for H.B.'s Ph.D. grant.

\begin{appendix}

\section{Correlations}\label{app:correlations}

Here, we give the dimensionless correlations obtained by solving equation (\ref{eq:evol_cor_adim}) when going from NESS to Equilibrium, and Equilibrium to NESS.

\subsection{NESS to Equilibrium}

The correlations are:
\begin{align}
\tilde c_{++} &=
-\frac{1}{2 A}	-\frac{f^2 u^2 e^{-B \tau}}{2 A B C} [B \sinh (\tau)+\cosh (\tau)],\\
\tilde c_{+-} &=-\frac{f^2 k^2 u^2 e^{-2 A \tau}}{2 A B C}-\frac{f^2 u^2 e^{2 \tau-2 A \tau}}{2 B C}-\frac{i f k u e^{\tau-2 A \tau}}{B C}+\frac{1}{2 A},
\end{align}
where $A=1+s^2$ ; $B=1+2s^2$ and $C=f^2 u^2+s^2+1$.
By symmetry, the other terms are $\tilde c_{--}=\tilde c_{++}$ and $\tilde c_{-+}=\tilde c_{+-}^*$.

\subsection{Equilibrium to NESS}

The correlations are:
\begin{align}
\tilde c_{++} &=
\frac{f^2 u^2 e^{-B \tau}}{2 A B C \Delta }
\left[B^2 \cosh \left(\sqrt{\Delta } \tau\right)+B \sqrt{\Delta } \sinh \left(\sqrt{\Delta } \tau\right)-4 C s^2\right] + \frac{1}{2 B C}-\frac{1}{B},\\
\tilde c_{+-} &=\frac{f u e^{-B \tau}}{2 A B C \Delta } \left[B \sqrt{\Delta } f u \sinh \left(\sqrt{\Delta } \tau\right) (B-2 i f s u)+B f u \cosh \left(\sqrt{\Delta } \tau\right) (\Delta -2 i B f s u)+2 i C s\right]\nonumber\\
&\qquad+\frac{1}{2 C}+\frac{i f s u}{B C},
\end{align}
where $A=1+s^2$ ; $B=1+2s^2$ ; $C=f^2 u^2+s^2+1$ and $\Delta= 1-4f^2s^2u^2$.
The other terms are $\tilde c_{--}=\tilde c_{++}$ and $\tilde c_{-+}=\tilde c_{+-}^*$.

\section{Regularity of the integrands from Equilibrium to NESS}\label{appendix:limit}

The integrands of the electrostatic and hydrodynamic corrections, Eqs.~(\ref{eq:integrand_etn_el}, \ref{eq:h}), seem to be singular when $1-4f^2s^2u^2$ approaches zero.
However, they are continuous and their limits are
\begin{align}
y &= \frac{2 f^3 u^4 e^{-\tau \left(\frac{1}{2 f^2 u^2}+1\right)}}{\left(2 f^2 u^2+1\right)^3 \left(4 f^2 u^2+1\right)} \nonumber\\
&\qquad \times\left[\left(2 f^2 \tau u^2+\tau\right)^2+\left(8 f^2 u^2+2\right) e^{\tau\left(\frac{\tau}{2 f^2 u^2}+1\right)}-8 f^2 u^2+4 \tau \left(2 f^4 u^4+f^2 u^2\right)-2\right],\\
h&=-\frac{4 f^2 u^2 \left(u^2-1\right)}{\left(2 f^2 u^2+1\right)^3 \left(4 f^2 u^2+1\right)} 
	\bigg[f^2 u^2 e^{\tau \left(-\frac{1}{2 f^2 u^2}-1\right)} \left(\left(2 f^2 \tau u^2+\tau\right)^2-2 \tau \left(2 f^2 u^2+1\right)-4 f^2 u^2\right)\nonumber\\
	&\qquad-2 \left(8 f^6 u^6+10 f^4 u^4+6 f^2 u^2+1\right)\bigg].
\end{align}

\end{appendix}




\bibliography{electro_transient.bib}

\begin{thebibliography}{10}
\providecommand{\url}[1]{\texttt{#1}}
\providecommand{\urlprefix}{URL }
\expandafter\ifx\csname urlstyle\endcsname\relax
  \providecommand{\doi}[1]{doi:\discretionary{}{}{}#1}\else
  \providecommand{\doi}{doi:\discretionary{}{}{}\begingroup
  \urlstyle{rm}\Url}\fi
\providecommand{\eprint}[2][]{\url{#2}}

\bibitem{Debye1923}
P.~Debye and E.~Hückel,
\newblock \emph{{Theory of electrolytes—part II: law of the limit of
  electrolytic conduction}},
\newblock {Physikalische Zeitschrift} \textbf{24}, 305 (1923).

\bibitem{Onsager1927}
L.~Onsager,
\newblock \emph{{Report on a revision of the conductivity theory}},
\newblock {Transactions of the Faraday Society} \textbf{23}, 341 (1927),
\newblock \doi{10.1039/tf9272300341}.

\bibitem{Onsager1957Wien}
L.~Onsager and S.~K. Kim,
\newblock \emph{{Wien Effect in Simple Strong Electrolytes}},
\newblock {The Journal of Physical Chemistry} \textbf{61}(2), 198 (1957),
\newblock \doi{10.1021/j150548a015}.

\bibitem{Dean1996}
D.~S. Dean,
\newblock \emph{{Langevin equation for the density of a system of interacting
  Langevin processes}},
\newblock {Journal of Physics A: Mathematical and General} \textbf{29}(24),
  L613 (1996),
\newblock \doi{10.1088/0305-4470/29/24/001}.

\bibitem{Demery2016Conductivity}
V.~Démery and D.~S. Dean,
\newblock \emph{{The conductivity of strong electrolytes from stochastic
  density functional theory}},
\newblock {Journal of Statistical Mechanics: Theory and Experiment}
  \textbf{2016}(2), 023106 (2016),
\newblock \doi{10.1088/1742-5468/2016/02/023106}.

\bibitem{Peraud2017}
J.-P. Péraud, A.~J. Nonaka, J.~B. Bell, A.~Donev and A.~L. Garcia,
\newblock \emph{{Fluctuation-enhanced electric conductivity in electrolyte
  solutions}},
\newblock {Proceedings of the National Academy of Sciences} \textbf{114}(41),
  10829 (2017),
\newblock \doi{10.1073/pnas.1714464114}.

\bibitem{Donev2019}
A.~Donev, A.~L. Garcia, J.-P. Péraud, A.~J. Nonaka and J.~B. Bell,
\newblock \emph{{Fluctuating Hydrodynamics and Debye-Hückel-Onsager Theory for
  Electrolytes}},
\newblock {Current Opinion in Electrochemistry} \textbf{13}, 1 (2019),
\newblock \doi{https://doi.org/10.1016/j.coelec.2018.09.004}.

\bibitem{Avni2022Conductivity}
Y.~Avni, R.~M. Adar, D.~Andelman and H.~Orland,
\newblock \emph{Conductivity of concentrated electrolytes},
\newblock Phys. Rev. Lett. \textbf{128}, 098002 (2022),
\newblock \doi{10.1103/PhysRevLett.128.098002}.

\bibitem{Avni2022Conductance}
Y.~Avni, D.~Andelman and H.~Orland,
\newblock \emph{Conductance of concentrated electrolytes: Multivalency and the
  wien effect},
\newblock The Journal of Chemical Physics \textbf{157}(15), 154502 (2022),
\newblock \doi{10.1063/5.0111645}.

\bibitem{Mahdisoltani2021}
S.~Mahdisoltani and R.~Golestanian,
\newblock \emph{{Long-Range Fluctuation-Induced Forces in Driven
  Electrolytes}},
\newblock {Phys. Rev. Lett.} \textbf{126}(15), 158002 (2021),
\newblock \doi{10.1103/PhysRevLett.126.158002}.

\bibitem{Kaiser2013}
V.~Kaiser, S.~T. Bramwell, P.~C.~W. Holdsworth and R.~Moessner,
\newblock \emph{{Onsager’s Wien effect on a lattice}},
\newblock {Nat Mater} \textbf{12}(11), 1033 (2013),
\newblock \doi{10.1038/nmat3729},
\newblock {Letter}.

\bibitem{Lesnicki2020}
D.~Lesnicki, C.~Y. Gao, B.~Rotenberg and D.~T. Limmer,
\newblock \emph{Field-dependent ionic conductivities from generalized
  fluctuation-dissipation relations},
\newblock Phys. Rev. Lett. \textbf{124}, 206001 (2020),
\newblock \doi{10.1103/PhysRevLett.124.206001}.

\bibitem{Lesnicki2021}
D.~Lesnicki, C.~Y. Gao, D.~T. Limmer and B.~Rotenberg,
\newblock \emph{On the molecular correlations that result in field-dependent
  conductivities in electrolyte solutions},
\newblock The Journal of Chemical Physics \textbf{155}(1), 014507 (2021),
\newblock \doi{10.1063/5.0052860}.

\bibitem{Richter2020}
{\L{}}.~Richter, P.~J. Żuk, P.~Szymczak, J.~Paczesny, K.~M. Bąk,
  T.~Szymborski, P.~Garstecki, H.~A. Stone, R.~Hołyst and C.~Drummond,
\newblock \emph{{Ions in an AC Electric Field: Strong Long-Range Repulsion
  between Oppositely Charged Surfaces}},
\newblock {Phys. Rev. Lett.} \textbf{125}(5), 056001 (2020),
\newblock \doi{10.1103/PhysRevLett.125.056001}.

\bibitem{Robin2021Modeling}
P.~Robin, N.~Kavokine and L.~Bocquet,
\newblock \emph{Modeling of emergent memory and voltage spiking in ionic
  transport through angstrom-scale slits},
\newblock Science \textbf{373}(6555), 687 (2021),
\newblock \doi{10.1126/science.abf7923}.

\bibitem{Robin2023Long-term}
P.~Robin, T.~Emmerich, A.~Ismail, A.~Niguès, Y.~You, G.-H. Nam, A.~Keerthi,
  A.~Siria, A.~K. Geim, B.~Radha and L.~Bocquet,
\newblock \emph{Long-term memory and synapse-like dynamics in two-dimensional
  nanofluidic channels},
\newblock Science \textbf{379}(6628), 161 (2023),
\newblock \doi{10.1126/science.adc9931}.

\bibitem{Kaiser2015AC}
V.~Kaiser, S.~T. Bramwell, P.~C.~W. Holdsworth and R.~Moessner,
\newblock \emph{ac wien effect in spin ice, manifest in nonlinear,
  nonequilibrium susceptibility},
\newblock Phys. Rev. Lett. \textbf{115}, 037201 (2015),
\newblock \doi{10.1103/PhysRevLett.115.037201}.

\bibitem{Mahdisoltani2021Transient}
S.~Mahdisoltani and R.~Golestanian,
\newblock \emph{Transient fluctuation-induced forces in driven electrolytes
  after an electric field quench},
\newblock New Journal of Physics \textbf{23}(7), 073034 (2021),
\newblock \doi{10.1088/1367-2630/ac0f1a}.

\bibitem{Dean2012}
D.~S. Dean, V.~Démery, V.~A. Parsegian and R.~Podgornik,
\newblock \emph{{Out-of-equilibrium relaxation of the thermal Casimir effect in
  a model polarizable material}},
\newblock {Phys. Rev. E} \textbf{85}(3), 031108 (2012),
\newblock \doi{10.1103/PhysRevE.85.031108}.

\bibitem{Dean2014}
D.~S. Dean and R.~Podgornik,
\newblock \emph{{Relaxation of the thermal Casimir force between net neutral
  plates containing Brownian charges}},
\newblock {Phys. Rev. E} \textbf{89}(3), 032117 (2014),
\newblock \doi{10.1103/PhysRevE.89.032117}.

\bibitem{Oksendal2000}
B.~Øksendal,
\newblock \emph{{Stochastic differential equations: an introduction with
  applications}},
\newblock {Springer}, {5th} edn.,
\newblock ISBN {9780387602431} (2000).

\bibitem{DeZarate2006Hydrodynamic}
J.~M.~O. De~Zarate and J.~V. Sengers,
\newblock \emph{Hydrodynamic fluctuations in fluids and fluid mixtures},
\newblock Elsevier (2006).

\bibitem{Kim2013}
S.~Kim and S.~J. Karrila,
\newblock \emph{{Microhydrodynamics: principles and selected applications}},
\newblock {Courier Corporation} (2013).

\bibitem{Brogioli2000Diffusive}
D.~Brogioli and A.~Vailati,
\newblock \emph{Diffusive mass transfer by nonequilibrium fluctuations: Fick's
  law revisited},
\newblock Phys. Rev. E \textbf{63}, 012105 (2000),
\newblock \doi{10.1103/PhysRevE.63.012105}.

\bibitem{Gardiner2009}
C.~Gardiner,
\newblock \emph{{Stochastic methods}},
\newblock {Springer Berlin} (2009).

\bibitem{Behr2019Solution}
M.~Behr, P.~Benner and J.~Heiland,
\newblock \emph{Solution formulas for differential {S}ylvester and {L}yapunov
  equations},
\newblock Calcolo \textbf{56}(4), 51 (2019),
\newblock \doi{10.1007/s10092-019-0348-x}.

\bibitem{Piessens2012Quadpack}
R.~Piessens, E.~de~Doncker-Kapenga, C.~W. {\"U}berhuber and D.~K. Kahaner,
\newblock \emph{Quadpack: a subroutine package for automatic integration},
  vol.~1,
\newblock Springer Science \& Business Media (2012).

\bibitem{Virtanen2020Scipy}
P.~Virtanen, R.~Gommers, T.~E. Oliphant, M.~Haberland, T.~Reddy, D.~Cournapeau,
  E.~Burovski, P.~Peterson, W.~Weckesser, J.~Bright, S.~J. {van der Walt},
  M.~Brett \emph{et~al.},
\newblock \emph{{{SciPy} 1.0: Fundamental Algorithms for Scientific Computing
  in Python}},
\newblock Nature Methods \textbf{17}, 261 (2020),
\newblock \doi{10.1038/s41592-019-0686-2}.

\bibitem{Poncet2017}
A.~Poncet, O.~Bénichou, V.~Démery and G.~Oshanin,
\newblock \emph{{Universal Long Ranged Correlations in Driven Binary
  Mixtures}},
\newblock {Phys. Rev. Lett.} \textbf{118}(11), 118002 (2017),
\newblock \doi{10.1103/PhysRevLett.118.118002}.

\end{thebibliography}

\nolinenumbers

\end{document}